\documentclass[%
 reprint,
superscriptaddress,
 amsmath,amssymb,
 aps,
]{revtex4-1}

\usepackage{graphicx}
\usepackage{dcolumn}
\usepackage{bm}

\usepackage{hyperref}

\usepackage[utf8]{inputenc}
\usepackage[T1]{fontenc}
\usepackage{mathptmx}
\usepackage{etoolbox}

\makeatletter
\def\@email#1#2{%
 \endgroup
 \patchcmd{\titleblock@produce}
  {\frontmatter@RRAPformat}
  {\frontmatter@RRAPformat{\produce@RRAP{*#1\href{mailto:#2}{#2}}}\frontmatter@RRAPformat}
  {}{}
}%
\makeatother

\usepackage{hyperref}

\usepackage{commath}
\usepackage{xcolor}
\usepackage{tabu}

\newcommand{\beq}{\begin{equation}}
\newcommand{\eeq}{\end{equation}}
\newcommand{\aver}[1]{\left\langle {#1}\right\rangle}
\newcommand{\bfx}{{\bf x}}
\newcommand{\bfr}{{\bf r}}
\newcommand{\nvec}{\hat{\bf n}}

\newcommand{\rvec}{\hat{\bf r}}
\newcommand{\bfR}{{\bf R}}
\newcommand{\bftheta}{{\bm \theta}}

\graphicspath{ {./} }
\DeclareGraphicsExtensions{.pdf}

\usepackage{subfig}
\begin{document}


\title{Inverse design of two-dimensional structure by self-assembly of patchy particles}

\author{Uyen Tu Lieu}
\email{uyen.lieu@aist.go.jp}
\affiliation{Mathematics for Advanced Materials-OIL, AIST, 2-1-1 Katahira, Aoba, 980-8577 Sendai, Japan}
\affiliation{Advanced Institute for Materials Research (AIMR), Tohoku University, 2-1-1 Katahira, Aoba, 980-8577 Sendai, Japan}

\author{Natsuhiko Yoshinaga}
\email{yoshinaga@tohoku.ac.jp} 
\affiliation{Mathematics for Advanced Materials-OIL, AIST, 2-1-1 Katahira, Aoba, 980-8577 Sendai, Japan}
\affiliation{Advanced Institute for Materials Research (AIMR), Tohoku University, 2-1-1 Katahira, Aoba, 980-8577 Sendai, Japan}
\date{\today}

\begin{abstract}
We propose an optimisation method for the inverse structural design of self-assembly of anisotropic patchy particles. The anisotropic interaction can be expressed by the spherical harmonics of the surface pattern on a patchy particle, and thus arbitrary symmetry of the patch can be treated. The pairwise interaction potential includes several to-be-optimised parameters, which are the coefficient of each term in the spherical harmonics. We use the optimisation method based on the relative entropy approach and generate structures by Brownian Dynamics simulations. Our method successfully estimates the parameters in the potential for the target structures, such as square lattice, kagome lattice, and dodecagonal quasicrystal.
\end{abstract}

\maketitle


\section{Introduction} 
The self-assembly of nano- and colloidal particles is a spontaneous organisation of the small particles into structures \cite{zhang_2017}. The self-assembly forming complex patterns is of great interest because of the promising applications in 
materials engineering, such as photonics, energy storage devices, tunable-rheology fluids \cite{zhang_2017,li_2020}. The special properties of these materials originate from the material structures and the physical-chemical properties of the components made of the materials \cite{boles_2016, sherman_2020}. Understand, control and predict self-assembly structure from given building blocks is a challenging goal in the field of soft matter. The self-assembly of complex structures, particularly open structures such as kagome lattice \cite{chen_2011}, honeycomb lattice \cite{goodrich_2021}, diamond structure \cite{he_2020}, icosahedral quasicrystal \cite{engel_2015}, however, has not fully understood yet. The main difficult issue is to determine building blocks which are capable of forming the desired structure, and the underlying kinetics and mechanism of the assemble process, i.e. the interaction of such building blocks and the conditions, e.g. temperature, dispersity will determine the outcome structure. The interaction of building blocks depends on the block and the solvent. The building block can be isotropic \cite{dotera_2014,engel_2015} 
 or anisotropic \cite{glotzer_2007}. In the later case, the building block is varied in shape \cite{damasceno_2012}, local sphere clusters \cite{marson_2019}, patchy particle \cite{zhang_2017,bianchi_2017}, etc. The patchy particle is often described as a spherical particle patterned with anisotropic surface, or attached with interacting patch on its surface \cite{zhang_2017,bianchi_2017,lieu_2020}. Recent developments in synthesis and fabrication techniques have enabled the realisation of those patchy particles \cite{chen_2011,vanoostrum_2015}. The assembly of patchy particles has led to the discovery of new order structures, phase diagram, and the extraction of some general features such as the formation of hierarchical assembly \cite{li_2016}.

Discovering the relation of particle design and self-assembled structure employs forward approach or inverse approach. In the conventional `forward' self-assembly approach, a specific potential or anisotropic particle is used to discover the assembled structure. Conversely, in `inverse' approach a class of computational technique is used to discover the suitable type of potential or anisotropic particle which is capable of forming a desired target structure \cite{jadrich_2017,ferguson_2018,sherman_2020}. For patchy particles, the infinite degrees of freedom of patchiness result in a great flexibility to achieve several complex structures. However, it also poses questions on the inverse problem: how to choose a suitable particle design for a specific structure among several possible particle designs. To determine suitable interactions for stabilising self-organisation, the optimasation techniques have been continuously developed for both isotropic interaction \cite{rechtsman_2005, rechtsman_2006,torquato_2009, ferguson_2018, pineros_2017, jadrich_2017, kumar_2019,goodrich_2021} and anisotropic interaction \cite{geng_2019,tracey_2019,romano_2020,whitelam_2020}.

In this study, we have proposed an inverse design strategy for the self-assembly of monodispersed patchy particles by Brownian dynamics simulation.  The algorithm is based on relative entropy method \cite{jadrich_2017} to optimise the yield of the target structure in the design parameter space. 
We apply the model to find suitable patchy particle designs for given target structures. The patchy particle in the study is a spherical particle whose surface pattern is based on spherical harmonics, thus any pattern can be described by a linear combination.\cite{lieu_2020} Such pattern consists of two type of regions which is an analogical to a polarised surface or even the surface charges of complex biological molecules \cite{daniel_2010,bianchi_2011,bianchi_2017}. Then one can set either repulsive or attractive interaction to the like/opposite patches, for example setting like patches repulsive and opposite patches attractive, which is similar to the charged interaction. The main feature of our patterned patchy particle system involves the complex interplay between the attractive and repulsive anisotropic interaction, and the capability of systematically exploiting the relation between the patchy particle symmetry and the self-assembly. Compared to the patchy particles with specific narrow attractive sites (sticky-patch model) where the patch number, patch arrangement and patch type are important features determining the assembled structure and properties, the particle in our approach are less involved with those factor. We expect that the experiment can be carried out more flexible by tuning the external parameters (solvent pH, salt concentration) \cite{vanoostrum_2015,bianchi_2017} and the fabrication required less strict conditions on such as a sophisticate geometry and chemical selectivity. 






 
\section{Methods} 
\label{methods}

\subsection{Pairwise interaction of patchy particles} 
In this study, the pattern of patchy particle $i$ is described by a spherical harmonic $Y_{lm}= Y_{lm}(\hat{\mathbf{x}})= \{ \mathbf{C}_{(i)}^{(l,m)} \} \odot \{ \hat{\mathbf{x}}^l  \}$, where the $l^{\rm{th}}$-rank irreducible tensor $ \{ \mathbf{C}_{(i)}^{(l,m)} \}$ includes the information of the local orientations of particle $i$ \cite{}, and $\odot$ is the $l$-fold contraction of two $l^{\rm{th}}$ rank tensors. Then for a pair of $Y_{lm}$ particles put at the distance $\mathbf r$, the anisotropic interaction $\Xi_{lm}$  is described as $\Xi_{lm}  \propto \{ \mathbf{C}_{(i)}^{(l,m)} \} \odot   \nabla_{\mathbf r}^{2l} \frac{1}{r}   \odot \{ \mathbf{C}_{(j)}^{(l,m)} \}$. For simplicity, we assume that the potential $u_{lm}$ for a pair of particles $Y_{lm}$ can be decomposed into a distance-dependent term and an orientation-dependent term as $u_{lm}=u_M(r)\Xi_{lm}(\hat{\mathbf{r}},\bm{\Omega})$. Here $u_M(r)$ is similar to the Morse potential, and $\Xi_{lm}(\hat{\mathbf{r}},\bm{\Omega})$ is dependent on the mutual orientation $\bm{\Omega}$ of the pair particle. We normalise  $\Xi_{lm}$ in $[-1,1]$.

\begin{figure}[ht]    	
   	 \centering
  	\includegraphics[width=0.45\textwidth, trim=0mm 55mm 0mm 45mm,clip]{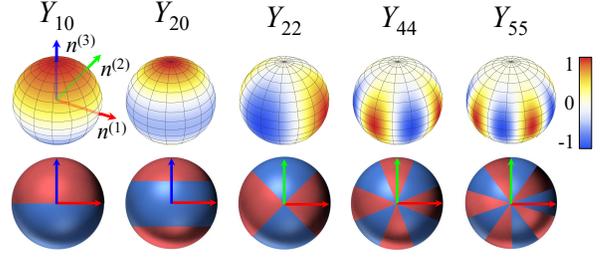}
   	\caption{The patchy particles described by spherical harmonic $Y_{lm}$ (top row) and the corresponding patchiness (middle and bottom rows). The orientation of a particle is characterised by the local orthonormal bases ($n^{(1)},n^{(2)},n^{(3)}$). The view views along the polar axis $n^{(1)}$ of each particle are shown in the bottom row.}
   	\label{fig:Y_lm}
	\end{figure}

 In the inverse problem, the to-be-optimised potential includes the interactions of the several types of patchy particles:
\begin{equation} \label{eq:u}
u=u_{WCA}(r) + \sum_{lm} \theta_{lm} u_{lm}(r,\bm{\Omega}),
\end{equation}
where the parameter $\theta_{lm}$ 
is the weight of each potential type to be optimised. For demonstration, the potential includes five candidates corresponding to five types of patchy particles $Y_{10}$, $Y_{20}$, $Y_{22}$, $Y_{44}$, $Y_{55}$, as shown in Fig. \ref{fig:Y_lm}. The particle $Y_{10}$ and $Y_{20}$ can be referred as Janus particle and triblock particle, respectively. They are axisymectric about the polar axis ($n^{(3)}$). The particle $Y_{22}$, $Y_{44}$ and $Y_{55}$ has twofold, fourfold, and fivefold symmetry in the equator plane spanned by ($n^{(1)}, n^{(2)}$). The patchiness of a particle can be determined by its positive (red) and negative (blue) pattern. In this model, $\theta>0$ means that different colour patch is attractive and similar one is repulsive, while $\theta<0$ means similar colour patch attractive and different is repulsive. The sign of $\theta$ implies different interaction and physics of the self-assembly, therefore it can be considered as a prior knowledge for the design of patchy particle. Since the sign of $\theta$ is often fixed by materials of the patches, we can set either $\theta \leq 0$ or $\theta \geq 0$. We may also set the sign unconstrained. The detail of the anisotropic function $\Xi_{lm}$, the Morse potential $u_M$, and the Week-Chandler-Anderson potential $u_{WCA}$ preventing the overlapping of particle and  can be found in Appendix. 
	
	The self-assembly of patchy particles is performed by Brownian dynamics\cite{allen_2017} under annealing. In each simulation, the spherical particles are confined to a flat plane while rotate freely in three dimensions. The periodic boundary condition is applied on the two-dimensional space. The number of particles in each iteration is $N=128$ for square lattice and kagome lattice, $N=240$ for dodecagonal quasicrystal. Initially the positions and orientations of particles are randomly distributed. Temperature is decresed from $T_{\text{max}}$ to $T_{\text{min}}$ with an interval of $\Delta T$. The values of $T_{\text{max}}$, $T_{\text{min}}$, $\Delta T$ are respectively $0.5,0.05, 0.01$ for square lattice and kagome lattice, and $1, 0.2, 0.0125$ for dodecagonal quasicrystal. 
 
\subsection{Optimisation scheme} 
Our optimisation process is based on the relative entropy approach \cite{jadrich_2017}. According to Jadrich et al. \cite{jadrich_2017}, a model for isotropic interaction is built as follows. A particle configuration $\bfR$ is described by a set of $N$ particles' positions $\bfR =[\bfr_1, \bfr_2, ...,\bfr_N]$. Let $P_{tgt}(\bfR)$ denote the probability distribution of a target structure, and $P(\bfR|\bftheta)$ denote the probability distribution for realising the configuration $\bfR$ given $\bftheta$. Here $P(\bfR|\bftheta)$ follows Boltzmann distribution $P(\bfR|\bftheta)=\frac{e^{-\beta U(\bfR| \bftheta)}}{Z(\bftheta)}$, in which the configurational partition function $Z(\bftheta)$ is a normalisation factor as $Z(\bftheta)=\int e^{-\beta U(\bfR| \bftheta)} d\bfR$, and $U(\bfR| \bftheta)$ is the tunable potential. One way to measure the distance between those two distributions is to use Kullback-Leibler divergence \cite{bishop_2006}. With the use of $\aver{f(\bfR)}_{P(\bfR|\bftheta)} = \int f(\bfR) P(\bfR |\bftheta) \, d\bfR $, the Kullback-Leibler divergence from $P(\bfR|\bftheta)$ to $P_{tgt}(\bfR)$ can be written as
	\beq 
	\begin{split} 
	D_{KL} (P_{tgt}||P) & =\int P_{tgt}(\bfR) \ln \frac{P_{tgt}(\bfR)}{P(\bfR|\bftheta)} \, d\bfR \\					
						 		& = \aver{\ln P_{tgt}(\bfR)}_{P_{tgt}(\bfR)} 	- 	\aver{\ln P(\bfR|\bftheta)}_{P_{tgt}(\bfR)}.
	\end{split} 
	\eeq    
In the optimisation scheme, $\bftheta$ is tuned to minimise $D_{KL}(P_{tgt}||P)$. The search for the local minima can be conducted by the gradient descent method. Then the next point of $\bftheta$ in the iteration is chosen by following the steepest descent so that:
	\beq
	\bftheta^{(k+1)}=\bftheta^{(k)}-\alpha \nabla_\bftheta D_{KL}(P_{tgt}||P),
	\label{eq:theta}
	\eeq
where $\alpha$ is the parameter controlling how far the point moves along the gradient descent curve. In this study we empirically set $\alpha=0.05$. Applying of Boltzmann distribution and partition function for the gradient term in Eq. \eqref{eq:theta}, we can rewrite the term as the following form 
	\beq \begin{array}{ll}
	\nabla_\bftheta D_{KL}(P_{tgt}||P) = &  \aver {\nabla_{\bftheta} \beta U(\bfR|\bftheta)}_{P_{tgt}(\bfR)} \\
												 &- \aver{\nabla_{\bftheta} \beta U(\bfR|\bftheta)}_{P(\bfR|\bftheta)} . 
	\end{array}
	\label{eq:nabla} 
	\eeq


In this study, the interaction of the particle is dependent on both the position and orientation of the particle, therefore, information on the orientation of the particle is required for the optimisation. When orientation of the target structure is available, Eq. \eqref{eq:nabla} is simply replaced by
\beq
\begin{array}{ll}
\nabla_{\theta} D_{KL}
\left(
P_{tgt} || P
\right)
=&	\langle
\nabla_{\theta} \beta U ({\bf R,n}| \theta)
\rangle_{P_{tgt} ({\bf R,n})}							\\
&-
\langle
\nabla_{\theta} \beta U ({\bf R,n}| \theta)
\rangle_{P({\bf R,n}|\theta)}.
\end{array}
\label{eq:nabla_ori} 
\eeq

As we discuss in Sec. \ref{target}, we also consider the target structure whose orientation is not measured. In this case, we treat the orientation of the target structure as a hidden variable. The optimisation is interpreted as minimisation between $P_{tgt} ({\bf R})$ and the probability distribution that is marginalised over the orientation, $P({\bf R}| \theta) = \int P({\bf R,n} | \theta ) P({\bf n})$. Then, we use, instead of Eq. \eqref{eq:nabla_ori},

\beq \begin{array}{ll}
\nabla_{\theta} D_{KL}
\left(
P_{tgt} || P
\right)
=&
\langle
\nabla_{\theta} \beta U ({\bf R,n}| \theta)
\rangle_{P_{tgt} ({\bf R}),P({\bf n})} 	\\
&-
\langle
\nabla_{\theta} \beta U ({\bf R,n}| \theta)
\rangle_{P({\bf R,n}|\theta)}
,
\end{array} \eeq
where $P({\bf n})$ is estimated from the generated structures for each iteration.

The weight $\theta$ affects the energy scale and how the patches on the particle interact. For each data set of optimisation, $\theta$ is limited to the range of $|\theta|=[0, 1.2]$. We also investigate the behaviour of $\theta$ when constraint on the sign $\bm{\theta}\geq \mathbf{0}$ or $\bm{\theta}\leq \mathbf{0}$ is applied.

\subsection{Target structures} 
\label{target}
In order to thoroughly evaluate the inverse design optimisation, the target structure is categorised into three groups in decreasing order of target's information: ground truth target, synthesised target, and minimal synthesised target. (i) The ground truth target is prepared by stabilising the particles using the potential in Eq. \eqref{eq:u} with  given values of $\bftheta$. Although the annealing scheme is chosen slow enough, the assembled structure still has defects and thermal fluctuations. The target includes all information about the positions and orientations of particles. It should be noticed that the calculation of energy of the assemblies requires the information of the position and the orientation (i.e. three local orthonormal bases $\nvec^{(1)}$, $\nvec^{(2)}$, $\nvec^{(3)}$, as shown in Fig. \ref{fig:Y_lm}) of the particles. (ii) The synthesised target is prepared based on the ground truth, however, the particle position is perfectly set according to the unit cell of a targeted lattice, the particle surface distance is set zero; then the orientations of the particles are set corresponding to an energy minimising structure. 
(iii) In the final case, only the position of the synthesised target is given to further check the capability of our proposed model. 
 
\begin{figure}[ht] 
   \centering
   
   \includegraphics[width=0.45\textwidth,trim=10mm 38mm 150mm 55mm,clip]{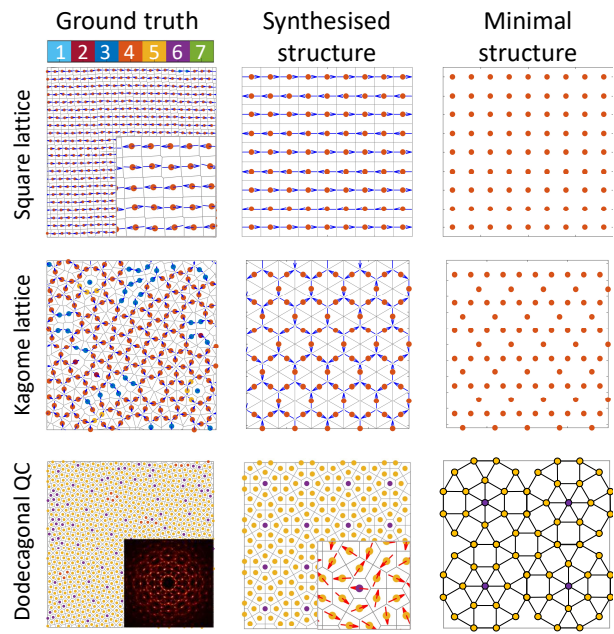}
   \caption{Illustration of target structures include square lattice, kagome lattice, dodecagonal quasicrystal considering ground truth (left), synthesised structure (middle), and minimal structrure (right). A Fourier transform of the ground truth dodecagonal quasicrystal with 12-fold symmetry is given inset. Other insets are enlarged structures. Particle colours are coded with number of nearest neighbours, the blue and red arrows are the orientations of the particles in the local direction $n^{(3)}$ and $n^{(1)}$ (Fig. \ref{fig:Y_lm}). The voronoi tesselation is also embeded.}
   \label{fig:target}
\end{figure}

In detail, we choose square lattice, kagome lattice and a dodecagonal quasicrystal as target structures (Fig. \ref{fig:target}). The square lattice is a basic structure. The kagome lattice is an open structure which is difficult to create via self-assembly and it is supposed to have novel optical and mechanical properties \cite{chen_2011}; dodecagonal quasicrystal is also difficult in terms of their aperiodic nature \cite{janssen_2018}. The ground truths of these structures, namely the square lattice, kagome lattice, dodecagonal quasicrystal can be obtained by setting the non-zero $\theta$ in Eq. \eqref{eq:u} as $\theta_{10}=1$, $\theta_{20}=-1$, $\theta_{55}=-1$, respectively. As for the synsthesised targets, particle positions and orientations of the square and the kagome lattice are perfectly assigned based on their unit cells. 
The synthesised dodecagonal quasicrystal target is constructed as an approximant crystal by suitable packed dodecagonal motif \cite{vanderlinden_2012}, however the orientation is too difficult to perfectly assign. We overcome this problem by numerically stabilising the approximant crystal under the potential type same to the one used for ground truth. Finally, the minimal targets, which have only information of particle position, are also considered. In this case, the orientation of target is a hidden parameter. However, because the orientation of the target is required for each iteration, we statistically interpolate the orientation fields from the generated structure and assign to the target. The algorithm of the interpolation of orientation can be found in Appendix \ref{app_assign}. 

\section{Results}
The optimisation starts with a random set of the to-be-optimised weight $\bftheta$, followed by a dynamical self-assemble process. Then the structural data at the final timestep, which is called generated structure, will be used for Eq. \eqref{eq:theta} to update the $\bftheta$ for the next iteration. The number of iterations is set around 20, 40 or hundreds. To compare the generated structures and the target, we generate the structure with the estimated parameters and then check the local structure, the number of neighbours and the radial distribution function. 
\subsection{Optimisation for ground truth and synthesised targets} 
\label{full}
\begin{figure*}[ht] 
   	\centering
	
 	\includegraphics[width=1\textwidth,trim=0mm 75mm 10mm 5mm,clip]{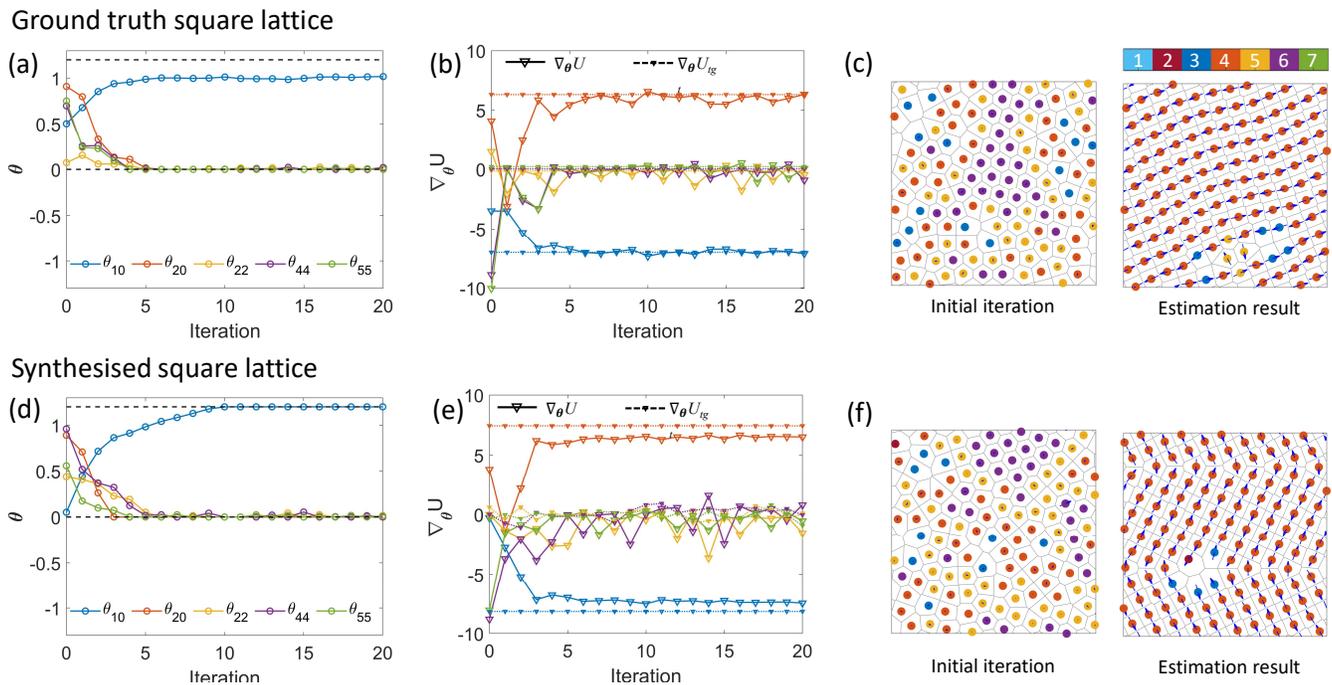}	
    	\caption{Optimisation for a ground truth square lattice (a,b,c), synthesised square lattice (d,e,f) from $Y_{10}$ particles. (a,d) The behaviour of $\bftheta$, a constraint on $\bftheta \geq 0$ is applied. (b,d) The behaviour of $\nabla_{\bftheta}U$ of generated structure (colour legends are similar to the $\bftheta$). (c,f) The snapshots are generated using the parameters at the initial condition and the estimated result. The colour of particle is the number of nearest neighbours of each particle. The blue arrows are the orientations of the particles in the local polar direction $n^{(3)}$ (Fig. \ref{fig:Y_lm}).}
    	\label{fig:square12all}
    	\end{figure*}

The performance of the optimisation method is confirmed by inverse design of the ground truth target structures. Such ground truth target structures are prepared from a specific $Y_{lm}$ potential as given in Eq. \eqref{eq:u}. The information of the ground truth target includes both positions and orientations of all particles. 
For the ground truth square lattice target assembled from $Y_{10}$ patterned particle, Fig. \ref{fig:square12all} shows the typical behaviours of the constrained $\bftheta \geq \bf{0}$. As shown in Fig. \ref{fig:square12all}(a,c), a solution containing the only non-zero $\theta_{10}\approx 1$ is quickly obtained, and the generated structure is identical to the ground truth target. We also prepare a synthesised lattice whose surface distance between particle is set zero, and orientation of polar basis $n^{(3)}$ is exact. Compared to the ground truth case, the behaviour of $\bftheta$ in Fig. \ref{fig:square12all}(d) is qualitatively similar. However the weight is $\theta_{10}=1.2$, which is the upper limit of $\bftheta$. The reason of this difference can be explained as follows: according to the estimation of $\bftheta$ given in Eq. \eqref{eq:theta} and \eqref{eq:nabla}, the terms $\nabla_{\bftheta}U$ of the generated structure and the target is distance-dependent. The distance between particles in the synthesised target is smaller than that of the ground truth (the first peak of radial distribution function for synthesised target and ground truth is $2a$ and $2.08a$ respectively; data are not shown). Since the interaction decays with distance, the term $\nabla_\bftheta U_{tgt}$ of synthesised target has greater magnitude as shown in Fig. \ref{fig:square12all}(b,e), leading to $\nabla_{\theta_{10}} U - \nabla_{\theta_{10}}U_{tgt} \neq 0$ for the synthesised lattice optimisation. As a result, $\theta_{10}$ keep increasing until it reaches the  upper limit. The result for the synthesised target suggests that the optimisation scheme can work when the distance between particle in the target is not strictly assigned, i.e. similar behaviour of $\bftheta$ is expected when the optimisation is conducted for a synthesised target whose distance between particle is larger or smaller. Additionally, it should be noted that in synthesised target, we assign only the polar basis (the $n^{(3)}$ in Fig.\ref{fig:Y_lm}) of particle while the information of azimuthal bases ($n^{(1)}$ and $n^{(2)}$ direction) are random; because the $Y_{10}$ particle is axisymetric, so the information of the polar basis is enough to attain a minimum energy structure. This result suggests that the model can work when the target contains partial information of the orientation. 

\begin{figure}[h] 
   	\centering  	
	
 	\includegraphics[width=0.45\textwidth,trim=0mm 73mm 150mm 5mm,clip]{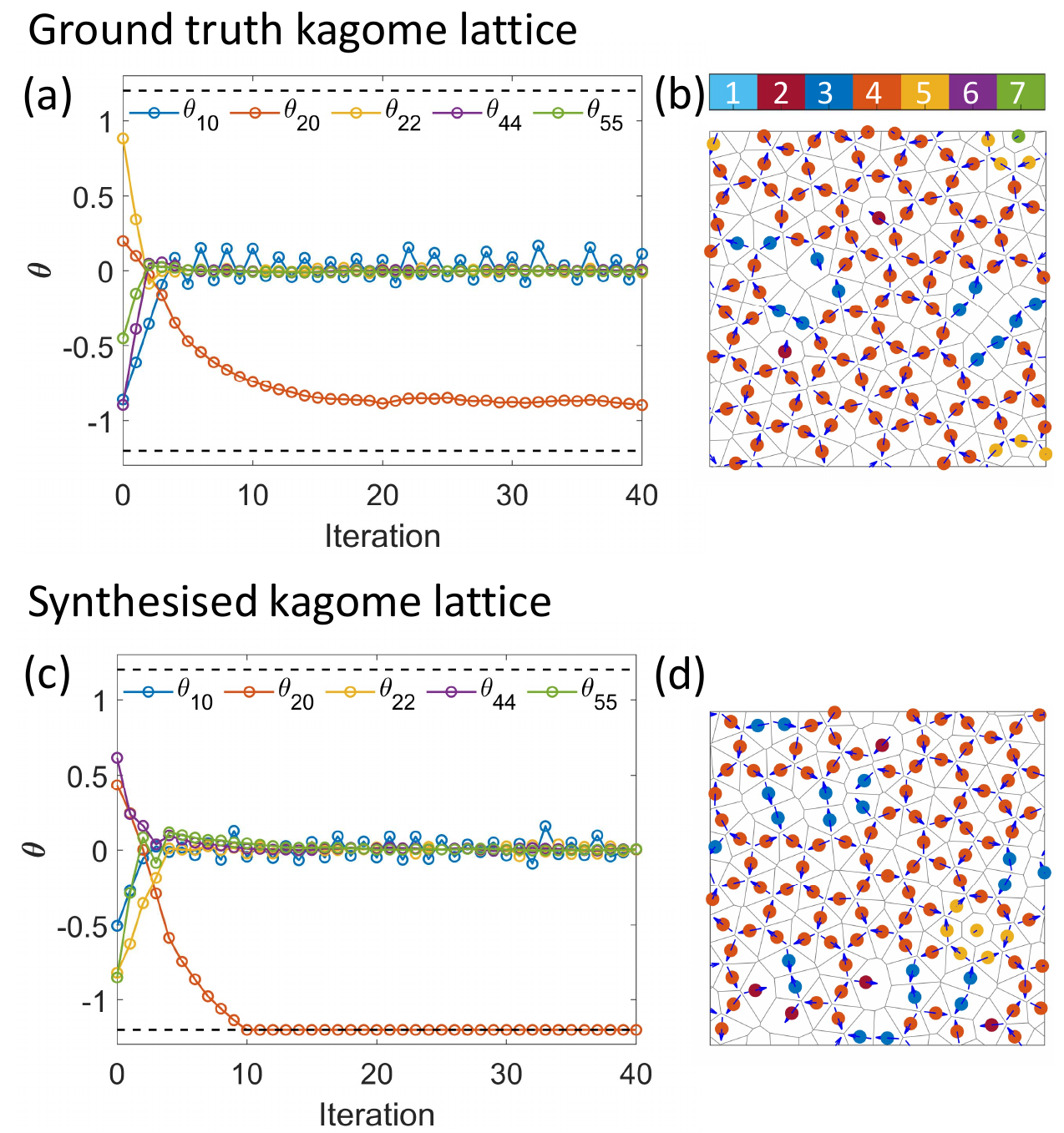}
    	\caption{The estimation of $\bftheta$ when the target is a ground truth kagome lattice (a-b), synthesized kagome lattice (c-d) assembled from $Y_{20}$ particles. The arbitrary sign constraint condition $\bftheta$ is applied. The snapshots are generated by estimated parameters. The blue arrows are the orientations of particles in the local polar direction.}
    	\label{fig:kagome12all}
	\end{figure}

	The result for kagome lattice is given in Fig. \ref{fig:kagome12all} when the sign of $\bftheta$ is arbitrarily assigned. The ground truth target is obtained by $\theta_{20}=-1$. The optimisation is capable of finding the correct solution whose nonzero value $\theta_{20} \approx -0.9$ for ground truth and $\theta_{20} = -1.2$ (lower limit) for synthesised target. The difference between the value of $\theta_{20}$ in such two cases is mostly due to the particle distance in the two targets, as mentioned in previous paragraph. One can see that the generated structure contains quite a lot of defects, perhaps this is the reason why $|\theta_{20}|$ for ground truth is slightly smaller than the expected value.

\begin{figure}[h] 
   	\centering
	
	\includegraphics[width=0.45\textwidth,trim=0mm 4mm 160mm 5mm,clip]{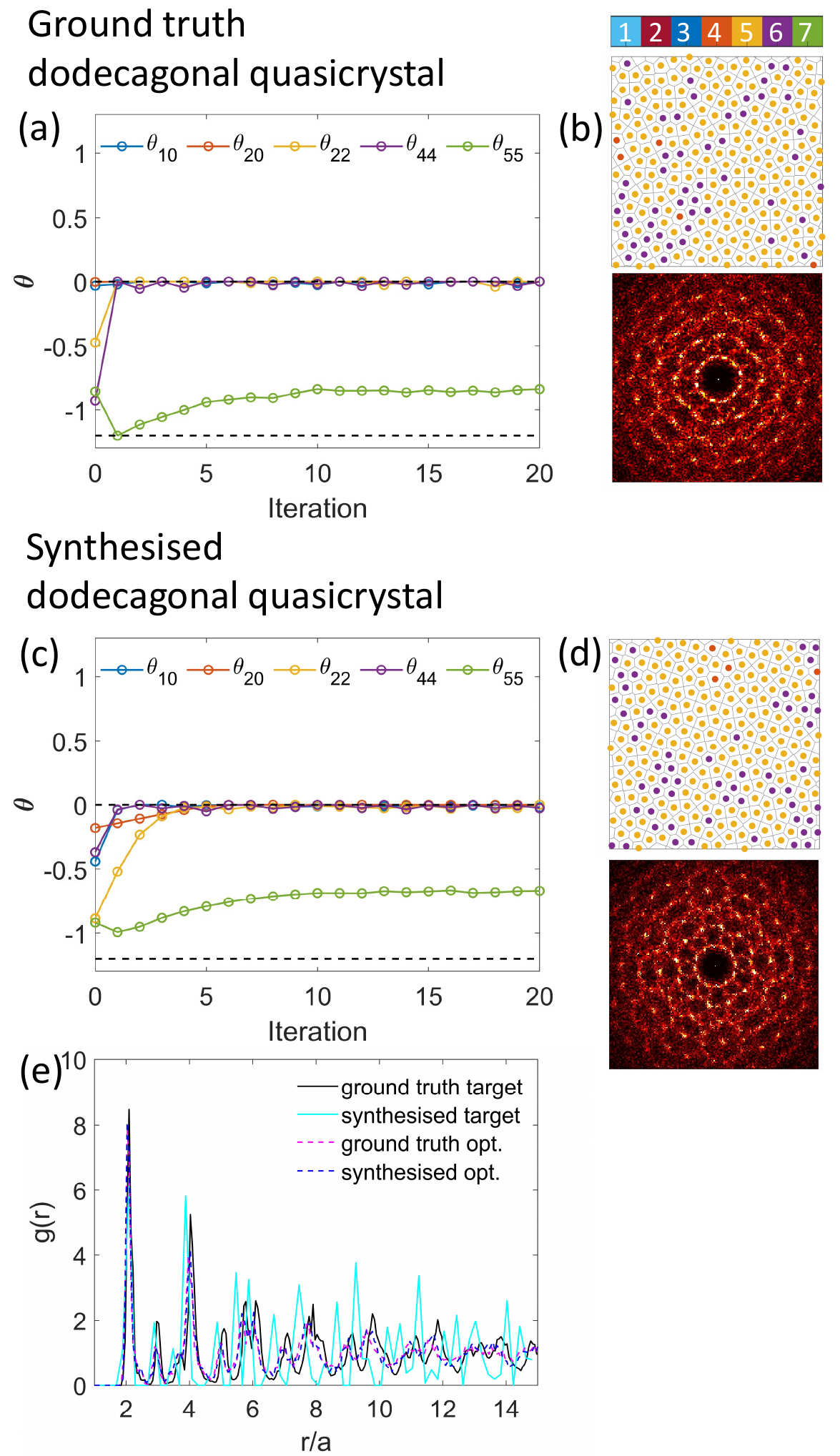} 
    	\caption{The estimation of $\bftheta$ when the target is a ground truth dodecagonal quasicrystal (a-b), synthesised approximant crystal (c-d) assembled from $Y_{55}$ particles. The negative signed constrain condition $\bftheta \leq 0$ is applied. Snapshots by estimated results and their Fourier transforms are included. (e) The comparison of the radial distribution functions for the ground truth target, synthesised target (Fig. \ref{fig:target}) and the snapshots in (b,d).}
    	\label{fig:dodeca12all}
    	\end{figure}

	The estimation for dodecagonal quasicrystal is given in Fig. \ref{fig:dodeca12all}. The ground truth is obtained by considering the potential with $\theta_{55}=-1$. For synthesised target, the same potential is used to numerically stabilise a position-fixed approximant crystal. Different from the synthesied target of square lattice and kagome lattice, we do this process to find the minimum energy state of the approximant because it is too difficult to assign the orientations to the particles. When $\bftheta \leq \bm{0}$, after a few iterations the optimisation model is capable of finding the suitable potential with $\theta_{55}\approx -0.9$ for ground truth and $\theta_{55}\approx -0.7$ for synthesised target. Although these values deviate from the expected value around $-1$, the convergence of $\bftheta$ behaves similarly and the only nonzero parameter for the estimation refers to the $Y_{55}$ patchy particle design. It is confirmed that the generated structures by those parameters have features of a dodecagonal quasicrystal, which are the 12-fold symmetry in their Fourier transforms in Fig. \ref{fig:dodeca12all}(b,d), and the similarity with the radial distribution of the ground truth target as in Fig. \ref{fig:dodeca12all}(e). Regarding the difference in estimation result, one possible reason is that the number of particles used in those simulation is quite small, therefore the periodic boundary condition can inhibit the growth of the quasicrystal and the relaxation of defects. For the synthesised target, the difference in the obtained value and expected value is more obvious. As mentioned previously, such effect is also caused by the particle distance in the generated structure and the target, which is also illustrated in Fig. \ref{fig:dodeca12all}(e).   
	
    	The defects in a generated structure have certain effect on the estimation of $\bftheta$. The defect is manifested in the form of grain boundary and point defect. For the square lattice case, the generated structure at the latter iterations contains a few defects and the obtained solution $\theta_{10} \approx 1$ ($\bftheta \geq \bf{0}$) is almost similar to the potential exclusively used to create the ground truth target. For the kagome lattice, it can be seen that the generated structure have quite a lot of defects. The generated structure of quasicrystal is not as clear as the ground truth, possibly because of the small number of particles. In these two case, the obtained solution ($\theta_{20}$ for kagome lattice and $\theta_{55}$ for dodecagonal quasicrystal) have smaller $|\theta|$ than expected. There is a trade-off between accuracy and computational cost, one may adjust the annealing setting of the simulation to reduce defect, however, the simulation cost increases. In this study, although the number of particles seems small and the annealing is quite fast, the results show sufficient ingredients for designing the particle.     	
  
 \subsection{Optimisation for target without information of particle orientation} 
For the minimal target, the knowledge on possible orientation of particle is unknown. Therefore we extract the orientation field of the generated structure and pass this information to the target. 
	
\begin{figure}[ht] 
   	\centering
    	
	 \includegraphics[width=0.45\textwidth,trim=0mm 12mm 150mm 5mm,clip]{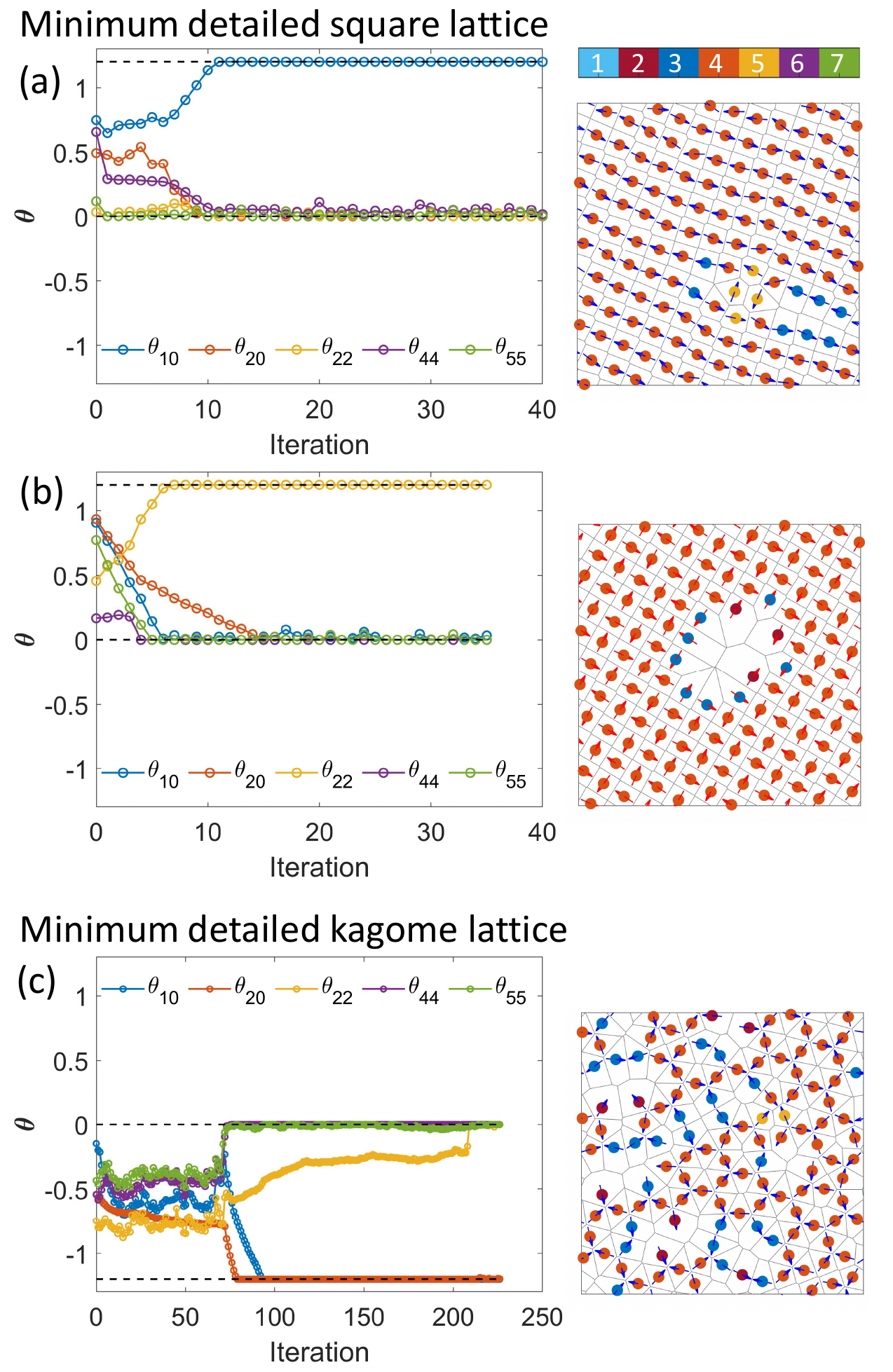}
    	\caption{The estimation of $\bftheta$ when target contains only particle position for square lattice (a,b) and kagome lattice (c). The snapshots by estimated result are included. The blue arrows and red arrows are the orientations of particles in the local $n^{(3)}$ and $n^{(1)}$ direction (Fig. \ref{fig:Y_lm}).} 
    	\label{fig:square3all}
    	\end{figure}

	Figure \ref{fig:square3all} illustrates the behaviours of $\bftheta$ and the possible solutions for square lattice and kagome lattice target. For the square lattice,  when the constraint $\bftheta \geq \bm{0}$ is applied (different patch attractive condition), we obtain two different sets of parameters. The nonzero parameters in these sets are $\theta_{10}=1.2$ and $\theta_{22}=1.2$, respectively. It suggests that aside from the $Y_{10}$ type as estimated in Fig. \ref{fig:square12all}(a), the other type of patchy particle, $Y_{22}$, can assemble into a square lattice. This patchy particle type composes of four patches around equator, and the patches with different color are attractive (see Fig. \ref{fig:Y_lm}). The capability of forming a square lattice for such kind of particle is intuitively understandable. 
	
	In the case of kagome lattice target with the constraint $\bftheta < \bf{0}$, as depicted Fig. \ref{fig:square3all}(c), we obtain the unique estimation in which there are two nonzero parameters as $\theta_{10}=\theta_{20}=-1.2$. The result suggests that the particle design is a combination of particle type $Y_{10}$ and $Y_{20}$. This result is different from the solution of  $Y_{20}$ found in previous sections (Fig. \ref{fig:kagome12all}). 
	
\begin{figure}[ht] 
	 \centering
 	  	
		\includegraphics[width=0.45\textwidth,trim=0mm 30mm 135mm 30mm,clip]{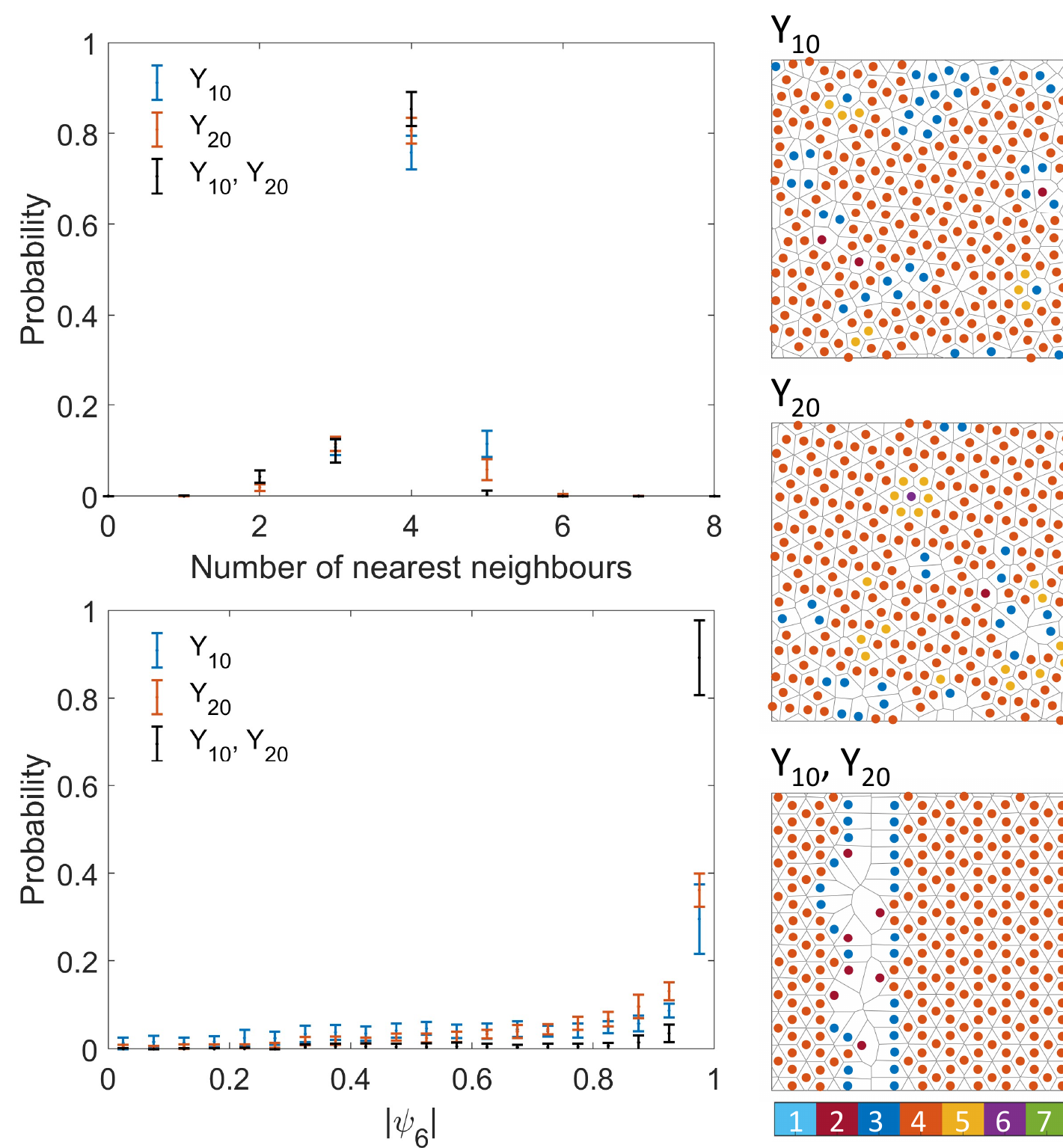}
    	\caption{Distribution of local number of neighbours and the six-fold bond-orientational order parameter $|\psi_6|$ for the self-assemblies of kagome lattice from patchy particle type $Y_{10}$, $Y_{20}$, and $Y_{10}+Y_{20}$. Representative snapshots for each particle type are included. The data point is shown by standard deviation from mean value based on about 30 independent configurations.}
    	\label{fig:kagome-errorbar}
	\end{figure}

	In order to evaluate the contribution of each type, we have independently conducted three groups of simulation using the particle patterned with $Y_{10}$, $Y_{20}$, and $Y_{10}+Y_{20}$. Figure \ref{fig:kagome-errorbar} shows a quantitative comparison of the structures obtained by the three particle types, by analysing the distribution of  the number of nearest neighbours and the six-fold bond-orientational order parameter \cite{nelson_2002} $|\psi_6|$. If a particle is a part of kagome lattice, it has four nearest neighbours and $|\psi_6| =1$. From the figure, for three types of patchy particle, the significant rate of $n_{nb}=4$ and $|\psi_6|\geq 0.7$ suggests that the structure is a kagome lattice \cite{chen_2011}. In the case $Y_{10}+Y_{20}$,  compared with the particles patterned only with $Y_{10}$ or with $Y_{20}$, the number of more-than-4-neighbour particles smaller while the number of less-than-4-neighbour particles larger, implying that the structure has more open local structure. Moreover, the system has more particles whose $\psi_6 > 0.9$ , suggesting the kagome lattice is clearer. This result reveals that the combination of $Y_{10}$ and $Y_{20}$ is able to create a much better kagome lattice than using only one of them. As depicted in Fig. \ref{fig:kagome-errorbar}, under the combined potential, the resulted kagome lattice has less defects, more open structure, and the shape of the unit cell is clearer than the case using the single component $Y_{10}$ or $Y_{20}$. A qualitatively similar phenomenon is observed for the enhanced self-assembly kagome lattice from tri-block particle whose the self-propel activity is added to the pole direction of tri-block particle \cite{mallory_2019}. The tri-block particle and the adding of self-propulsion can be respectively referred to the $Y_{20}$ and $Y_{10}$ in our study. Although the nature of enhancement by the self-propulsion \cite{mallory_2019} is different from the patchy particle $Y_{10}$, they are similar in terms of symmetry of the interaction between the particles. These results suggest that our study is capable of finding new designs for patchy particle so that the self-assembly is enhanced.

\section{Discussion and Conclusion}

In summary, we have developed a relative entropy-based method for inverse structural design of patchy particle for a given target structure. The type of a patchy particle is described by the spherical harmonic. The pairwise interaction potential includes several to-be-optimised parameters, which are the coefficient of each term in the spherical harmonics. We successfully estimate patches necessary to reproduce the targets such as two-dimensional square lattice, kagome lattice and dodecagonal quasicrystal. The method also works for hidden information for the square and kagome lattices, i.e. the target contains only the minimal information of the position of particles. The estimation is dependent on the type of a target structure, namely, prior knowledge of the structure that we want to reproduce. The prior knowledge of the range of the parameters is also crucial for the estimation. Here, we discuss those effects, as well as the possible extension of the model.  

\subsection{Effect of prior knowledge of target structure }
We have considered three kinds of target structure: the ground truth target comprises both the particle position and orientation, the synthesised target has its particle position set in perfect order and orientation obtained via minimum energy, and the minimal target has only information of the particle position. For the periodic target like square lattice and kagome lattice, the optimisation model is capable of finding the suitable solution design so that the desired targets can be obtained. Regarding the performance of the estimation for three types of target, it shows that as the more information on the target  is provided, the faster the optimisation process is. For example in the case of square lattice target (Fig. \ref{fig:square12all} and Fig. \ref{fig:square3all}), for the ground truth and synthesised targets, the stable and converging solution can be consistently obtained after just a few iterations, whereas the minimal target requires more iterations. The minimum target does not have data on the orientation, so the optimisation involves searching on a higher dimension space. As a result, it is more difficult to estimate the parameters. We also observe that the occurrence for finding the correct solution of minimum target is often lower than the other target. Among many independent estimation processes for each square lattice target, the solutions for ground truth and synthesised ones are all consistent and can reproduce a square lattice structure, while for the minimum target cases about 50$\%$ of the estimations can. However, such disadvantage is compensated by the fact that more possible solutions and ``better'' solutions can be found. In detail, when the minimum targets are considered, we have found two possible particle designs for the square lattice, three for the kagome lattice in which a combination design improves the lattice. On that account, one can consider different levels of prior knowledge for target structure. Here is an example for the self-assembly of kagome lattice. From the lattice, one can think that there can be at least two ways to design the particle's patches and the corresponding orientation of those particle in the lattice. Intuitively, the design based on narrow sticky patch consists of four small patches located on the same plane at alternating interval of 60$^{\circ}$ and 120$^{\circ}$ so that they fit the four contact points \cite{doppelbauer_2010, antlanger_2011, whitelam_2016,ranguelov_2019}. However, self-assembly of those particles shows that a rhombus lattice is more favourable and formed instead of kagome one \cite{doppelbauer_2010, antlanger_2011, whitelam_2016,ranguelov_2019}. The kagome lattice structure is formed only when constraint on patch selectivity is added \cite{whitelam_2016,ranguelov_2019}. Another solution is the axisymmetric particle, then the orientation of particle can be inferred, which is similar to the one in synthesised target. Using such kind of information for the target will result in the finding of $Y_{20}$ particle design, in which two opposite identical caps are attractive while the equatorial part is repulsive \cite{chen_2011,romano_2011a, mallory_2019}. 

In the case of minimum target dodecagonal quasicrystal, the optimisition is not able to find the expected solution. Aside from the difficulty caused by the hidden orientation as discussed above, there are several factors hindering the process. The target does not have information of particle orientation, so it is necessary to assign the orientation from the generated structure to the target. In the generated structure, the effect of thermal fluctuation is inevitable. Such fluctuation is also captured and passed to the target. In other words, simply imagine that the number of patches on the particle $Y_{55}$ of quasicrystal is large, the size of the patch is thus small, and then the fluctuation may lead to a considerable effect. So the effect of this fluctuation on the estimation outcome is more severe than that of the square lattice and kagome lattice. Another factor is that compared to the formation of the aperiodic lattice like dodecagonal quasicrystal is difficult and thus requires more investigation on the suitable temperature range, annealing rate, as well as the influence of the other potential components.  


\subsection{Effect of constraint parameters }
\begin{figure}[ht] 
   	\centering
	
 	\includegraphics[width=0.45\textwidth,trim=0mm 74mm 155mm 5mm,clip]{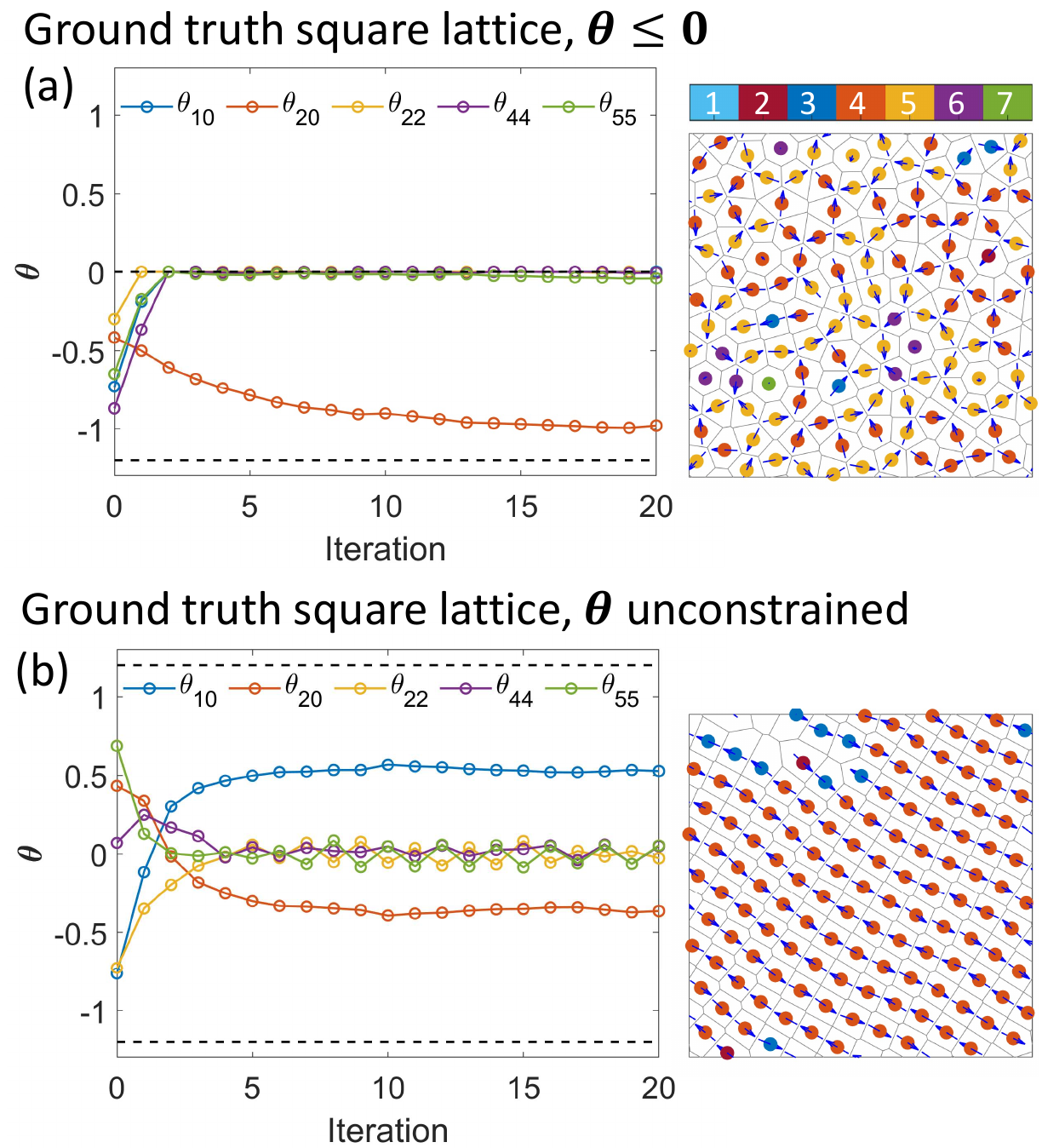}	
    	\caption{Estimation of $\bftheta$ at different constraint conditions when target is a ground truth square lattice assembled from $Y_{10}$ particles: (a) $\bftheta < 0$ , (b) unconstrained sign $\bftheta $. The snapshots are at the last iterations. }
    	\label{fig:Gsquare}
    	\end{figure}

In this study, the sign of $\theta$ involves the way the patch interacts, thus the potential type and fabrication technique. The patchy particle has two kind of regions (red and blue) on its surface, and different-kind region is set attractive for $\theta > 0$, while same-kind region is set attractive for $\theta < 0$. It should be noted that the self-assembly is completely different when the sign changes. Applying constraint on the sign of theta also affect the optimisation result. We analyse this issue via the optimisation of the ground truth square lattice in three cases: $\bftheta \geq \bf{0}$, $\bftheta \leq \bf{0}$, and unconstrained sign. When $\bftheta \geq \bf{0}$, the unique solution is $\theta_{10}\approx 1$ and the generated structure is similar to the target (Fig. \ref{fig:square12all}). When $\bftheta < \bf{0}$, as shown in Fig. \ref{fig:Gsquare}, the only solution is $\theta_{20}\approx -1$, however the structure consists of many 4-, 5-, 6-nearest neighbour particles and is quite irregular. When the sign of $\bftheta$ is not constrained, we are able to obtain a lattice similar to the target in terms of both position and orientation, however the solution is a combination of $\theta_{10}\approx 0.5$ and $\theta_{20}\approx -0.5$. The contribution of $\theta_{10}>0$ to the formation of the square lattice is well confirmed. It facilitates the head to tail alignment of the $Y_{10}$ particles, and the side-by-side or parallel alignment. The $\theta_{20} < 0$ corresponds to the tri-block particle similar to the one for kagome lattice. The rise of $\theta_{20} < 0$ in this case enhances the head-to-tail configuration, and possibly also enhances the parallel alignment because the particle with $\theta_{20}<0$ (i.e. the equatorial parts are attractive) is more energetically favourable. It is interesting that using $\theta_{20}<0$ alone creates a undesired structure, but a combination with another design $\theta_{10}>0$ can create the desired structure.   About the choice of sign-constraint, it depends on the situation, for example when the interaction of the red and blue patch is already set, then the constraint is a must. Setting either positive or negative $\bftheta$ often requires less number of iterations because the design parameter space is reduced. The consideration of unconstrained $\theta$ may spend more computational cost, but the estimation can lead to richer possible outcomes of patchy particle design.  

 As mentioned earlier, the absolute value of the coefficient $\theta$ refers to the contribution of spherical harmonics to patchy particle design. In the case of square lattice, for example, the estimated result is $\theta_{10}=1.0$ or $\theta_{10}=1.2$ depending on the target. The difference of these values does not matter much, because they both lead to the target structure, and suggests the same particle design. 

\subsection{Effect of the number of parameters}
\begin{figure}[ht] 
   	\centering
 	\includegraphics[width=0.45\textwidth,trim=0mm 90mm 153mm 5mm,clip]{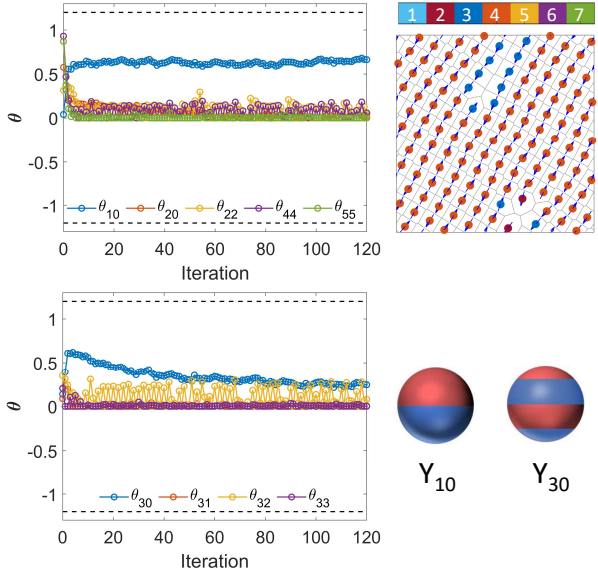}
    	\caption{9-parameter optimisation for ground truth square lattice. Data are split into 2 groups for better presentation.The generated structure by estimation result, and the estimation of particle patchiness are also given.}
    	\label{fig:10prm}
    	\end{figure}

In the scope of this study, the pool includes five types of patchy particles. It is straightforward to increase the number of parameters. We have increased the number of parameters to 9 by adding four more types of patchy particle $Y_{30}$, $Y_{31}$, $Y_{32}$, $Y_{33}$ and performed inverse optimisation for a ground truth square lattice. Regarding the computation cost, the time for each iteration increases by around 30$\%$ compared to the 5-parameter case. The computational cost scales almost linearly in the number of parameters. This is appealing compare with the conventional grid search, in which the cost scales exponentially. As shown in Fig. \ref{fig:10prm}, the estimation result of $\bftheta$ is unique and consist of two nonzero parameters $\theta_{10}\approx 0.65$ and  $\theta_{30} \approx 0.25$. The estimation is comparable to the 5-parameter case in Fig. \ref{fig:square12all}(a). The rise of $\theta_{30}$ is understandable because the $Y_{30}$ has similar symmetry to the $Y_{10}$, which promote the head-tail configuration. There is some fluctuation for $\theta_{32}$, but such fluctuation does not affect the generated structure. Although the inclusion of many parameters makes the behaviours of the self-assembly and the estimation more complicated, it is expected to reveal more combinations and suggestions on the design of patchy particle.      

\begin{acknowledgments}
The authors acknowledge the support from JSPS KAKENHI Grant number JP20K14437 to U.T.L., and JP20K03874 and JP20H05259 to N.Y.
\end{acknowledgments}

\section*{Data Availability}
The data that support the findings of this study are available from the corresponding author upon reasonable request.

\appendix
\section{Anisotropic interaction of the patchy particle}
The anisotropic interaction is calculated based on the mutual orientation of a pair of particle $i$ and $j$. Let the $\nvec^{(m)}_i$, $\nvec^{(m)}_j$, $m=1,2,3$ are local bases of particle $i$ and $j$ (Fig. \ref{fig:Y_lm}), and $\rvec$ is the unit distance vector between particle center. The anisotropic interaction $\Xi_{lm}  \propto \{ \mathbf{C}_{(i)}^{(l,m)} \} \odot   \nabla_{\mathbf r}^{2l} \frac{1}{r}   \odot \{ \mathbf{C}_{(j)}^{(l,m)} \}$ estimates the angular dependent of a pair of particles $Y_{lm}$ as $\Xi_{lm}  \propto \{ \nvec_0^{l-m} \nvec_+^{m}\}_{(i)} \odot   \{\rvec^{2l} \}   \odot  \{ \nvec_0^{l-m} \nvec_+^{m}\}_{(j)} $, where $\nvec_0=\nvec^{(3)}$ and $\nvec_+=\frac{1}{\sqrt{2}}(\nvec^{(1)} + i \nvec^{(2)})$. For example,  $\Xi_{10}  \propto \{ \nvec_0 \}_{(i)} \odot   \{\rvec \rvec \}   \odot  \{ \nvec_0  \}_{(j)} $ for a pair of particles $Y_{10}$, and $\Xi_{20}  \propto \{ \nvec_0  \nvec_0\}_{(i)} \odot   \{\rvec \rvec \rvec \rvec \}   \odot  \{ \nvec_0 \nvec_0 \}_{(j)} $ for a pair of particles $Y_{20}$.

\section{Potential}
The detail of the isotropic Week-Chandler-Anderson potential $u_{WCA}$ preventing the overlapping of particle, and the Morse potential $u_M$ in Eq.\eqref{eq:u} is given as 
	\begin{equation}
	u_{WCA}= 
  	\begin{cases}
  	4\varepsilon \left[ (\frac{2a}{r})^{12}- (\frac{2a}{r})^{6}+\frac{1}{4} \right]	, & r\leq 2a\sqrt[6]{2} \\
  	0 												    							, & r > 2a\sqrt[6]{2}
  	\end{cases} 
	\end{equation}

	\begin{equation}
	u_M= \varepsilon M_d \left \{ \left[ 1-e^{\left( -\frac{r-r_{eq}}{M_r} \right)} \right]^2 -1 \right \} 
	\end{equation}
where $\mathbf{r}=\mathbf{r}^{ij}=\bm{r}^j-\bm{r}^i$ is the distance vector between particle centres, $r=\left|\mathbf{r}\right|$, and $\hat{\mathbf{r}}=\mathbf{r}/r$, $\varepsilon$ is the potential well depth, $r_{eq}$ is the Morse potential equilibrium position ($r_{eq}=1.878a$), $M_d$,  $M_r$ is the Morse potential depth and range respectively ($M_d=3.761a$, $M_r=0.213a$ for short-range potential and $M_d=2.294a$, $M_r=a$ for long-range potential \cite{delacruz-araujo_2016}). 

\section{Local structure analysis}
For each particle $j$, the number of nearest neighbours $n_{\rm{nb}}$ is determined by counting the number of particle $k$ satisfying $r_{jk} \leq 2.5a$ where $a$ is the radius of the particle.

The n-fold bond-orientational order parameter \cite{nelson_2002} $\psi_n$ of particle $j$ is calculated by $\psi_n(j)=\frac{1}{n_{\rm{nb}}}\sum_{k=1}^{n_{\rm{nb}}} e^{in\varphi_{jk}}$, where $\varphi_{jk}$ is the angle between particle $j$ and its neighbouring particle $k$. The $|\psi_n|$ value characterises the local degree of the regular $n$-fold order around a particle; for example, the hexagonal lattice has $|\psi_6|=1$, kagome lattice also has $|\psi_6|=1$ but the number of nearest neighbours is $n_{\rm{nb}}=4$.

\section{Assignment of orientation} 
\label{app_assign}
For a given target, information of the target is required. Such information is often the positions of particles, or the density field. In the case of patchy particle, the orientation indicating the patches must be known. In reality the orientation may not be measured. In this case, the orientation must be treated as a hidden variable. Our approach is that the orientation of the target is interpolated from that of the generated structure during each iteration.

From the position and orientation of the particles in generated structure, we aim to (i) compute the local orientation field of the generated structure, and (ii) statistically estimate the orientations of particles in targeted structure. From a given generated structure consisting particles of the position $\{\bfx_1, \bfx_2,...\}$ and orientation $\{\nvec_1, \nvec_2,...\}$, where $\bfx \in R^2$ and $\nvec \in R^3$ in this study, the orientation field $\nvec(\bfx)$ is locally estimated. Consider the particle $i$, the neighbour $j$ can be determined if the distance $||\bfx_j - \bfx_i|| \leq r_{nb}$, and let $m(i)$ be the number of neighbours of particle $i$. Then $\nvec(\bfx)$ is taken from the probability distribution $P(\bfx_j,\nvec_j)$ under fixed position and orientation of particle $i$, i.e. translate and rotate the $j$th particles so that $\bfx_i=\bm{0}$ and $\nvec_i$ is on some axis. This is written as $\bfx_j \mapsto \bfx_{ji}=\bfx_j- \bfx_i$ and $\nvec_j \mapsto R(\nvec_j|\nvec_i)$. The position of $\bfx_j$ is implicitly rotated. Apply kernel density estimator for arbitrary position $\bfx$ in the system, the density of position of $j$ and orientation of $j$ is determined as 
	\beq
	g_x(\bfx)=\frac{1}{N_i}\sum_{i} \frac{1}{m(i)} \sum_j G(\bfx-\bfx_{ji})=\aver{G(\bfx-\bfx_{ji})}_{ij} 
	\label{eq:kernelGx}
	\eeq 
	
	\beq
	g_n(\nvec,\bfx)=\aver{G(\bfx-\bfx_{ji}) \, G(\nvec-R(\nvec(\bfx_j)|\nvec_i))}_{ij}
	\label{eq:kernelGn}
	\eeq
where $\bfx_{ji}=\bfx_j-\bfx_i$ (after the transformation), $N$ is the number of particle $i \in [1,N]$, $G(x)$ is Gaussian kernel (kernel with the shape of a Gaussian curve) defined and normalised as $G(x)=\frac{1}{\sqrt{2\pi}\sigma } e^{-\frac{x^2}{2\sigma^2}}$, and $\aver{.}_{ij}$ is mean over $i$ and $j$. Here Gaussian kernel is applied for both position $\bfx$ and orientation $\nvec$ in Eq. \eqref{eq:kernelGx} and \eqref{eq:kernelGn}. For the $n$-dimension data, we apply a one dimensional Gaussian curve sequentially in the $n$ dimensions. The equations \eqref{eq:kernelGx} and \eqref{eq:kernelGn} includes the density distribution of the neighbours orientations and position around a given particle. Then the orientation of an arbitrary neighbour $j'$, positioned at $\bfx_{j'}$ around the centre particle $i'$, denoted as $\nvec_{j'}(\bfx_{j'}  | (i',j'))$ can be interpolated from the density 
	\beq
	g(\nvec|\bfx)=\frac {g_n(\nvec,\bfx)} {g_x(\bfx)}
	\eeq

\textbf{Computational consideration:} 

(1) About $R(\nvec_j|\nvec_i)$: The orientation of particle consists of three local orthonormal bases $\nvec^{(m)}$, $m=1,2,3$ in Cartesian coordinates. In principal, when translate and rotate the $j$th particle so that $\bfx_i=\bm{0}$ and $\nvec_i$ is on some axis, one can choose $\nvec_i^{(1)}$, $\nvec_i^{(2)}$, $\nvec_i^{(3)}$ align with $Ox$, $Oy$, $Oz$, respectively; as a result, the position of particle $j$th after the transformation $\bfx_j \mapsto \bfx_{ji}=\bfx_j- \bfx_i$ is in three-dimensional space, which makes the calculation of rotational transformation, Eq. \eqref{eq:kernelGx}, \eqref{eq:kernelGn} and  $\nvec_{j'} (\bfx_{j'}  | (i',j')) $ more complex. In this study, since $\bfx$ is in $xy$ plane,  we simplify those calculations by choosing the two bases almost lying on the $xy$ plane, then align their $xy$-projection with the axis $Ox, Oy$. As a result, the position of particle is always on $xy$ plane, and the rotation $\nvec_j \mapsto R(\nvec_j|\nvec_{i\text{, projected}})$ is two-dimensional.

(2) For the Gaussian kernel $G(x)$ and $G(n)$, the position of the particle is in polar coordinate $x=[0, 2\pi ]$, while the orientation $\nvec$ is in Cartesian coordinate $n = [-1,1]$. The width of Gaussian kernel is chosen so that the full width at half maximum is around 30-50 times smaller than the range of variable. We choose $\sigma_x = 0.1$, $\sigma_n=0.02$. 

(3) Only particles in the 1st shell (i.e. around the first peak of pair distribution function) are considered. 

(4) Assign the orientation: As mentioned above, two local bases of particle orientation $\nvec^{(m)}$, $\nvec^{(n)}$ are independently estimated, then the 3rd basis is determined $\nvec^{(l)}= \nvec^{(m)} \times \nvec^{(n)}$. Since the three bases are required to be orthogonal, singular value decomposition is applied to find the nearest orthogonal matrix for the three bases. During the assignment, the constraint $\nvec^{(m)}.\nvec^{(n)}< 0.6$ is employed to reduce the deformation of the bases before and after applying SVD. (Note: to find the orthogonal basis, another way is just fix $\nvec^{(m)}$ and rotate $\nvec^{(n)}$ to the nearest orthogonal vector). 

(5) The assignment of orientation for target structure can be performed in two ways: global or local. (i) In global assignment, from an initial pair $(i,j)$, the calculation is then propagated to all particles of target. Orientation for $j$th particle is calculate $m(j)$ times, and we have $m(j)$ configuration for target. Note that we can not take the mean of orientation (i.e. the average $ 
\nvec_2(\bfx_j)=\frac{1}{m(j)} \sum_{i}{\nvec_2(\bfx_j|(i,j))} $ due to its circular quantity nature. (ii) In local assignment, the calculation is only for neighbours $j$ of particle $i$.\\

Assignment of orientations for all particles is not chosen, because the further the neighbours are, the more fluctuating their orientations are. 

\bibliography{MyLibrary}
\bibliographystyle{unsrt}
\end{document}